\documentclass[%
preprint,
longbibliography,
amsmath,amssymb,
aps,
prf,
]{revtex4-2}

\usepackage{ae,lmodern} %Choix de la police de caractères
\usepackage[english]{babel} %Gère les conventions typographiques à la langue en option, et défini les noms ``Chpter'', ``Section'', ``Table of contents'', ... dans la langue désirée.
\usepackage[utf8]{inputenc} %Indique le codage du fichier. UTF8 permet l'utilisation des caractères accentués dans le fichier source
\usepackage[T1]{fontenc} %Codage des caractères en T1. Gestion améliorée des caractères accentués dans le fichier de sortie.
\usepackage{microtype} %Apporte des raffinements typographiques supplémentaires (en particulier l’ajustement de l’espacement permettant une meilleure coupure des mots).

\usepackage{array}
\usepackage{makecell}

\usepackage{hyperref} %permet la navigation au sein d’un document PDF lors de l'utilisation des références.
\hypersetup{                    % parametrage des hyperliens
    colorlinks=true,                % colorise les liens
    breaklinks=true,                % permet les retours à la ligne pour les liens trop longs
    urlcolor= blue,                 % couleur des hyperliens
    linkcolor= blue,                % couleur des liens internes aux documents (index, figures, tableaux, equations,...)
    citecolor= blue,                % couleur des liens vers les references bibliographiques
	}
\usepackage{graphicx} %Pour utiliser includegraphics
\usepackage{subfig} %Pour les figures multiples
\usepackage{tikz} %Package de dessin. Utilisé ici pour mettre les styles de lignes dans les caption des figures.
\tikzstyle{long dashdotted}=[dash pattern=on 6pt off 3pt on \the\pgflinewidth off 3pt] %Style de ligne supplémentaire
\graphicspath{{Figures/}} %Dossier dans lequel se trouve les figures

\definecolor{C0}{RGB}{31,119,180}
\definecolor{C1}{RGB}{255,127,14}
\definecolor{C2}{RGB}{44,160,44}
\definecolor{C3}{RGB}{214,39,40}
\definecolor{C4}{RGB}{148,103,189}

\usepackage{amsmath, bm} %Contient des environnements supplémentaires comme align,
\usepackage{natbib}
\begin{document}
%\preprint{APS/123-QED}

\title{A modified kinetic theory for frictional-collisional bedload transport valid from dense to dilute regime.}
\author{R\'emi Chassagne}
\email{remi.chassagne@univ-grenoble-alpes.fr}

\affiliation{%
	Univ. Grenoble Alpes, LEGI, CNRS UMR 5519 - Grenoble, France
}%

\author{Julien Chauchat}
\affiliation{%
	Univ. Grenoble Alpes, LEGI, CNRS UMR 5519 - Grenoble, France
}%

\author{Cyrille Bonamy}
\affiliation{%
	Univ. Grenoble Alpes, LEGI, CNRS UMR 5519 - Grenoble, France
}%
\date{\today}

\begin{abstract}	
	Modelling sediment transport is still a challenging problem and is of major importance for the study of particulate geophysical flows. In this work, the modelling of sediment transport in the collisional regime is investigated with a focus on the continuum modelling of the granular flow. For this purpose, a frictional-collisional approach, combining a Coulomb model with the kinetic theory of granular gases, is developed. The methodology is based on a comparison with coupled fluid-discrete simulations, that classical kinetic theory model fails to reproduce. In order to improve the continuum model, the fluctuating energy balance is computed in the discrete simulations and systematically compared with the kinetic theory closure laws. Inter-particle friction is shown to affect the radial distribution function and to increase the energy dissipation, in agreement with previous observations in the literature. Due to saltating particles, whose motion can not be captured by the kinetic theory, departure from the viscosity and diffusivity laws are observed in the dilute part of the granular flow. Finally, the quadratic nature of the drag force is shown to increase the granular fluctuating energy dissipation. Based on these observations, modifications of the kinetic theory closure laws are proposed. The modified model reproduces perfectly the discrete simulations in the entire depth structure of the granular flow. These modifications are shown to impact the rheological properties of the flow and they make it possible to recover the $\mu(I)$ rheology in the dense regime.
\end{abstract}

\maketitle

\section{Introduction}

Sediment transport is one of the main processes that shapes river beds and coastal regions and is of major importance for their management and interactions with human infrastructures. Modelling the transport of particles by a flowing fluid is therefore a great challenge for the prediction of the morphological evolution of our environment. According to \citet{berzi2013}, sediment transport regimes may be differentiated using the bed slope $\alpha$ and the Shields number $\theta=\tau_b/(\Delta \rho g d)$, where $\tau_b$ is the fluid bed shear stress, $\Delta \rho$ is the density difference between particles and fluid, $g$ is the gravity acceleration and $d$ is the particle diameter. At bed slopes above approximately $5^\circ$  debris flows are observed for all Shields numbers. Below this value, increasing the Shields number from just above the critical value, sediment transport takes place as ordinary bedload corresponding to a single layer of grains moving on top of the fixed sediment bed. For $\theta \ge 0.2$ a transition is observed from collisional sediment transport to collisional transport with turbulent suspension at larger Shields numbers. While ordinary bedload may be modelled using statistical approaches (e.g. \cite{einstein1937, ancey2020a}) collisional transport and turbulent suspension are usually modelled in the continuum framework. The challenge in these approaches lies in the complexity of the particle-particle and turbulence-particle interactions modelling. Over the last two decades, extensive research has been carried out on this problem. Most of the time, Reynolds-averaged turbulence modelling have been used such as mixing length model (e.g. \cite{jenkins1998, revil-baudard2013}) or two-equations models such as $k-\epsilon$ or  $k-\omega$ (e.g. \cite{hsu2004a, lee2016, chauchat2017, gonzalez-ondina2018}). More recently, Large Eddy Simulation has been used to model turbulence in these two-phase flows (e.g. \cite{cheng2018, mathieu2021}). Concerning the granular stress modelling, two approaches have been used, $\mu(I)$ rheology (e.g. \cite{revil-baudard2013, lee2016, chauchat2017}) and Kinetic Theory of granular flows (e.g. \cite{jenkins1998, hsu2004, berzi2011, cheng2018}). This work investigates the collisional transport regime with a particular focus on the modelling of the granular part of the flow.\\

In order to model granular flows continuously, two approaches are generally considered depending on the granular flow regime. Dense dry granular flows can be described with the phenomenological $\mu(I)$ rheology \citep{gdrmidi2004, dacruz2005, pouliquen2006}, which relates the stress state, represented by the shear to normal stress ratio or effective friction coefficient $\mu=\tau^p/P^p$, to the kinetic state of the granular flow, represented by the inertial number $I = d|\dot{\gamma}|/\sqrt{P^p/\rho^p}$, where $d$ is the particle diameter, $\rho^p$ the particle density and $\dot{\gamma}$ is the granular velocity shear rate. This phenomenological law, derived by fitting data, gives predictive results for dense granular flows but it fails in the dilute regime. In the bedload configuration, \citet{maurin2016} studied the rheology of monodisperse beds with coupled fluid-Discrete Element Method (DEM). Despite the presence of water, they showed that the dry inertial number is still the controlling parameter. They found the $\mu(I)$ rheology to be valid in the dense part of the granular flow and extended the classical laws for dry granular flows to the bedload case as follows
\begin{equation}
\mu(I) = \mu_s + \dfrac{\mu_2-\mu_s}{I_0/I+1},
\label{eq::muI}
\end{equation}
\begin{equation}
\phi(I) = \dfrac{\phi_{c}}{1+bI},
\label{eq::phiI}
\end{equation}
with $\mu_s=0.35$ is the static friction coefficient below which no motion is \textit{a priori} possible, $\mu_2=0.97$, $I_0=0.69$, $\phi_{c} = 0.61$ and $b=0.31$. These laws persist for inertial number of the order of unity but depart from data at higher inertial numbers in the dilute regime.

Dilute granular flows are generally described with the Kinetic Theory (KT) of granular gases, which is based on a statistical description of the granular flow and binary collisions between particles \citep{chapman1970}. It relies on the Enskog-Boltzmann equation for the velocity distribution function and differs from the classical KT of molecular gases by taking into account the finite size of particles and energy dissipation at contacts due to inelasticity. It provides hydrodynamics conservation equations for mass, momentum and granular temperature. The granular temperature is defined as the averaged granular fluctuating kinetic energy and is not at all related to thermal temperature. These equations need closures for the stress tensor, granular temperature diffusivity and rate of dissipation. In the framework of the KT they can be expressed as complex integrals of the particle distribution function (solution of the Enskog-Boltzmann equation) and of the radial distribution function $g_0$, which characterizes the degree of spatial correlation between two particles that are going to collide. Under certain assumptions, analytical expressions for the coefficients may be obtained by application of the Chapmann-Enskog method \citep{chapman1970}, leading to different models. The main difficulty is to express the particle distribution function. The simplest model were not solving the Enskog-Boltzmann equation and assumed to be at equilibrium (no temporal nor spatial gradients) leading to the classical Maxwell distribution function \citep{savage1981, jenkins1983}. \citet{lun1984} considered a small perturbation around the Maxwell distribution and obtained constitutive relations for the coefficients that are valid for weakly dissipative granular flows, i.e. $1-e^2 <<1$ with $e$ the restitution coefficient. More recently, performing expansion to first order in spatial gradients, Garzo and Dufty \cite{garzo1999} derived relations for the whole range of restitution coefficient. Considering the large spatial variations of all quantities in the bedload configuration, i.e. exponential velocity profile, this latter model is certainly the most relevant one for the present study.

Considering bedload, and environmental flows over an erodible bed in general, the difficulty arises from the coexistence of all granular flow regimes with transition from a dilute granular flow at the bed surface to dense and quasi-static regime inside the bed.  In terms of continuum modeling strategy, two options are possible:  extend the $\mu(I)$ rheology to the dilute regime or extend the KT to the dense regime. In the dilute regime, non-local effects are important features of the flow and it seems therefore hardly feasible to model all regimes with a local rheology. Consequently the second approach is adopted hereafter. \\

There are various reasons to observe departure from the KT even in the dilute part of the flow. First because natural particles are frictional while the classical KT models such as the one of \citet{garzo1999} have been derived for frictionless particles. Interparticle friction makes it possible to transfer translational kinetic energy into rotational kinetic energy during collisions and add another source of energy dissipation by sliding motion. There have been some attempts to generalize KT to frictional particles including the resolution of rotational kinetic energy and a tangential restitution coefficient \citep{rao2008}. This however makes models even more complex and some authors proposed a more pragmatic approach in which the restitution coefficient is empirically reduced to account for interparticle friction dissipation \citep{jenkins2002, chialvo2013}. 

At the top of the sediment bed particles are transported by saltation with ballistic trajectories. Their motion is controlled by gravity and fluid drag force and is therefore out of the scope of the KT \citep{berzi2020}. Therefore, the KT is not expected to reproduce bedload in the very dilute part of the flow. There have been few attemps to model saltation in the continuous framework \cite{jenkins2018} and this is still a challenge for the modelling of geophysical particulate flows.

Finally, the KT is not expected to predict the behavior of the granular flow in the dense regime. This is because the motion of particles starts to be correlated and the molecular chaos assumption breaks down while particles are more concentrated. To model dense flows, \citet{jenkins2006, jenkins2007} and  \citet{berzi2011} proposed the introduction of a correlation length in the dissipation term of the temperature equation, showing good results when comparing with DEM shear flow simulations \citep{berzi2014, berzi2015}. In addition to these considerations, the origin of stresses changes when going into the bed. While stresses are due to fluctuating motions and very short collisions in the dilute regime, force chains start to emerge with long lasting contacts in the dense part of the bed leading to elastic stresses that can not be captured by the KT. \citet{johnson1987} proposed that the elastic stresses follow a Coulomb-like behavior and assumed that they can simply be summed up with the kinetic stresses to compute the total stresses. With this approach, \citet{berzi2020} were able to reproduce DEM simulations of dry granular flows on an erodible bed.\\

The literature review presented above highlights the duality between the $\mu(I)$ rheology, valid in the dense regime, and the KT \textit{a priori} valid in the dilute regime. A two-fluid model based on the KT that would be able to reproduce quantitatively the entire depth structure of the bedload layer would represent a major contribution for the sediment transport community. In particular, reproducing the $\mu(I)$ rheology with a KT based model is still an open question. In order to focus on the granular flow modelling and on the particle-particle interactions, collisional transport only is considered without any turbulent suspension and fluid-particle interactions will be reduced to the simplest ones, \textit{i.e.} buoyancy and drag forces. 

In this context, the aim of the present work is to propose a two-fluid model for bedload transport using the KT for granular flows and based on the frictional-collisional approach first proposed by \citet{johnson1987}. The model is implemented in SedFoam \citep{chauchat2017}, which is an open-source three dimensional (3D) solver based on the OpenFOAM toolbox. Following the same approach as \citet{chialvo2013} in simple shear cell granular flows, comparison with coupled fluid-DEM simulations will be performed. Keeping in mind the difficulties highlighted previously, empirical modifications to the classical KT model of \citet{garzo1999} will be proposed in order to reproduce quantitatively the discrete simulations. 

First the two phase flow model (section~\ref{sec::twophasemod}) and the coupled fluid-DEM model (section~\ref{sec::dem}) are presented. The results of both models will be compared and some modifications to the \citet{garzo1999} model will be proposed in section~\ref{sec::results} and implemented in SedFoam to quantitatively reproduce the discrete simulations. Finally, results are discussed in section~\ref{sec::discussion}.

\section{Two-phase flow models for bedload transport} \label{sec::twophasemod}

\subsection{Two-fluid model}

The 3D two-fluid model for sediment transport described in \citet{chauchat2017}, based on the equations derived by \citet{jackson2000}, is used in this contribution. Although it has been written in 3D, the model will only be used in 1D, with all variables depending only on the vertical axis $z$. Herein, only the simplified 1D equations will be presented \citep{chauchat2018}, and the interested reader is referred to \citet{chauchat2017} for more details on the 3D model. The release of the code used in this work can be found in \citet{bonamy2021}. In the following, overscript $^f$ (resp. $^p$) denotes for fluid (resp. particle) phase quantity. The mass and momentum conservation equations aims at solving the fluid and solid phase volume fraction, respectively $\epsilon$ and $\phi = 1-\epsilon$, as well as the Favre-averaged fluid and particle phase velocities, respectively $\bm{u}^f = (u_x^f, u_z^f)$ and $\bm{v}^p = (v_x^p, v_z^p)$. The Favre-averaged operators are detailed in appendix~\ref{sec:app1}.

\subsubsection{Mass and momentum conservation equations}

The mass conservation equations of the fluid and particle phase reduce to
\begin{equation}
\dfrac{\partial \rho^f \epsilon }{\partial t} + \dfrac{\partial \rho^f \epsilon u_z^f}{\partial z} = 0,
\label{eq::massFluid}
\end{equation}
\begin{equation}
\dfrac{\partial \rho^p \phi }{\partial t} + \dfrac{\partial \rho^p \phi v_z^p}{\partial z} = 0,
\label{eq::massPart}
\end{equation}
where $\rho^f$ and $\rho^p$ are respectively the fluid and particle density. The momentum conservation equations in the streamwise direction are written as
\begin{equation}
\dfrac{\partial \epsilon \rho^f u_x^f}{\partial t} + \dfrac{\partial u_z^f \epsilon \rho^f u_x^f}{\partial z} = \epsilon \rho^f g \sin(\alpha) + \dfrac{\partial \tau_{xz}^f}{\partial z} - nf_{Dx},
\label{eq::momxFluid}
\end{equation} 
\begin{equation}
\dfrac{\partial \phi \rho^p v_x^p}{\partial t} + \dfrac{\partial v_z^p  \phi \rho^p v_x^p}{\partial z} = \phi \rho^p g \sin(\alpha) + \dfrac{\partial \tau_{xz}^p}{\partial z} + nf_{Dx},
\label{eq::momxPart}
\end{equation} 
and in the wall-normal direction 
\begin{equation}
\dfrac{\partial \epsilon \rho^f u_z^f}{\partial t} + \dfrac{\partial u_z^f \epsilon \rho^f u_z^f}{\partial z} =  - \epsilon \rho^f g \cos(\alpha) -\epsilon \dfrac{\partial p^f}{\partial z} - nf_{Dz},
\label{eq::momzFluid}
\end{equation} 
\begin{equation}
\dfrac{\partial \phi \rho^p v_z^p}{\partial t} + \dfrac{\partial v_z^p \phi \rho^p v_z^p}{\partial z} = -\phi \rho^p g \cos(\alpha) - \dfrac{\partial p^p}{\partial z} - \phi \dfrac{\partial p^f}{\partial z} + nf_{Dz},
\label{eq::momzPart}
\end{equation} 
where $p^f$ (resp. $p^p$) is the fluid (resp. particle) pressure, $\tau_{xz}^f$ (resp. $\tau_{xz}^p$) is the fluid (resp. particle) shear stress, $g=9.81$ $m^2.s^{-1}$ is the gravity acceleration, $\alpha$ is the bed slope angle and $n\bm{f_{D}} = n(f_{Dx}, f_{Dz})$ is the drag force between the fluid and particle phase.

\subsubsection{Fluid phase closures}

To solve the fluid equations~\eqref{eq::massFluid}, \eqref{eq::momxFluid} and \eqref{eq::momzFluid}, it is necessary to have closures for the fluid shear stress $\tau_{xz}^f$ and drag force $n\bm{f_D}$. The fluid shear stress is the sum of a viscous shear stress and turbulent shear stress. The latter stress, also called Reynolds shear stress, is computed based on an eddy viscosity concept and the total shear stress expresses
\begin{equation}
\tau_{xz}^f = \rho^f\epsilon(\nu^f + \nu_t)\dfrac{\partial u_{x}^f}{\partial z},
\end{equation}
where $\nu^f$ is the fluid kinematic viscosity. The turbulent viscosity $\nu_t$ follows a mixing length approach that depends on the integral of the solid volume fraction to account for the presence of the particles \citep{li1995}
\begin{equation}
\nu_t = l_m^2 | \dfrac{\partial u_{x}^f}{\partial z}|, \quad l_m(z) = \kappa \int_{0}^{z}1-\dfrac{\phi(\xi)}{\phi_{max}}d\xi,
\end{equation}
with $\kappa = 0.41$  the Von-Karman constant. This simple formulation of mixing length extends
the \citet{prandtl1926} law for flow inside and over a mobile sediment bed, and has been widely used for sheet-flow and bedload applications \citep{li1995, dong1999, revil-baudard2013, maurin2015}.

The drag force is computed as
\begin{equation}
n\bm{f_{D}} = \phi(1-\phi)K\left(\bm{u}^f - \bm{u}^p \right), \quad K = 0.75 C_D \dfrac{\rho^f}{d}||\bm{u}^f - \bm{u}^p||\left(1-\phi\right)^{-\zeta-1}.
\label{eq::dragForce}
\end{equation}
The $\left(1-\phi\right)^{-\zeta-1}$ term accounts for hindrance effects due to the collective presence of particles and the exponent is fixed to $\zeta = 3.1$ \citep{jenkins1998}. The drag coefficient $C_D$ is computed following \citet{dallavalle1943}
\begin{equation}
C_D = C_D^{\infty} + \dfrac{24.4}{Re_p}, \quad Re_p = \dfrac{||\bm{u}^f - \bm{u}^p||d}{\nu^f},
\label{eq::dragCoef}
\end{equation}
where $C_D^{\infty} = 0.4$ is the value of the drag coefficient in the limit of infinite particle Reynolds number.

\subsubsection{Particle phase closures}

As already pointed out in the introduction, granular stresses have two physical origins. First, particles endure elastic stresses, resulting from enduring contacts, which do not depend on shear-rate. These elastic stresses appear at high volume fractions and lead to a frictional Coulomb-like behavior of the granular bed. Secondly, particles also endure kinetic stresses resulting from transfer of momentum due to fluctuating motions and at collisions between particles. In order to account for both these stresses in our model, the approach of \citet{johnson1987}, which consists in adding both contributions to the total granular stress, is adopted,
\begin{align}
\tau_{xz}^p &= \tau_{xz}^{el} + \tau_{xz}^{kin},\\
p^p &= p^{el} + p^{kin}.
\end{align}

\paragraph{Elastic stresses}

Models for elastic stresses originate from soil mechanics or poro-elasticity and are, for most of them, empirical. In this paper, the model proposed by \citet{johnson1987} is used,
\begin{equation}
p^{el} = \left\{
\begin{array}{lr}
0, & \phi < \phi_{rlp} ,\\
P_0\dfrac{(\phi-\phi_{rlp})^3}{(\phi_{max}-\phi)^5}, & \text{otherwise}
\end{array}
\right.
\label{eq::pel}
\end{equation}
with $P_0 = 0.05$ ($kg.m^{-1}.s^{-2}$), $\phi_{rlp}=0.57$ the random loose packing fraction. The shear stress is then computed following a Coulomb model,
\begin{equation}
\tau^{el} = \mu_s p^{el},
\label{eq::tauel}
\end{equation}
where $\mu_s=0.35$ is the effective static friction coefficient computed by \citet{maurin2016} with discrete element simulations of bedload transport. Note that the elastic pressure only exists for volume fraction larger than $\phi_{rlp}$ and the elastic stresses are therefore restricted to the very dense part of the granular flow. \\

\paragraph{Kinetic stresses}

As already mentioned, the kinetic stresses are computed in the framework of the KT. One needs to solve an energy balance equation for the particle phase in addition to the set of equations~\eqref{eq::massPart}, \eqref{eq::momxPart} and \eqref{eq::momzPart}. For multiphase flows, the rate of change of particle fluctuating kinetic energy can be written as \citep{ding1990},
\begin{equation}
\dfrac{3}{2}\left(\dfrac{\partial \rho^p \phi T}{\partial t} + \dfrac{\partial u_z^p \rho^p \phi T}{\partial z} \right) = \tau_{xz}^{kin}\dfrac{\partial v_x^p}{\partial z} + \dfrac{\partial q}{\partial z} + \Gamma + J_{int},
\label{eq::TPart} 
\end{equation} 
where $T=1/3\left< v_i^{p\prime}v_i^{p\prime}\right>^s$ is the granular temperature and $\left<\cdot\right>^s$ denotes for the particle phase average. The first term in the Right Hand Side (RHS) is a production term of temperature by the work of the mean granular flow. The second one in the RHS is a diffusion term with $q$ the granular temperature flux,
\begin{equation}
q = -\kappa \dfrac{\partial T}{\partial z}.
\label{eq::heatFlux}
\end{equation}
The third term in the RHS is dissipation of granular temperature due to contact inelasticity. The last term is specific to multiphase flows and represents the work done by the drag force due to fluctuating motion. In order to close the granular temperature equation, the \citet{garzo1999} model is used. It has been derived for inelastic hard spheres to first order in the spatial gradients. The transport coefficients are expressed as a function of the granular temperature as
\begin{align}
\label{eq::pkin}
p^{kin} &= \rho^p F_1(\phi) T, \\
\label{eq::etakin}
\eta^{kin} &= \rho^p d F_2(\phi) \sqrt{T},\\
\label{eq::kappakin}
\kappa &= \rho^p d F_3(\phi) \sqrt{T},\\
\label{eq::gammakin}
\Gamma &= \rho^p/d F_4(\phi) T^{3/2},
\end{align}
where $\eta^{kin} = \tau_{xz}^{kin}/|\dot{\gamma}|$ is the viscosity associated with the kinetic stress. The expressions of $F_1$, $F_2$, $F_3$ and $F_4$ are reported in the left column of table~\ref{tab::kin_th} and depend on the restitution coefficient $e$. Note that the dimensionless viscosity and diffusivity are expressed as the sum of a kinetic (due to fluctuations) collisional and bulk contributions.

\begin{table}
	\resizebox{\textwidth}{!}{$
	\begin{array}{rcl|rcl}
	\multicolumn{3}{c|}{\text{\cite{garzo1999} model}} & \multicolumn{3}{c}{\text{Proposed model}}\\
	\hline
	F_1(\phi) &=& \phi\left(1 +2(1+e)\phi g_0(\phi) \right) &  \multicolumn{3}{c}{\text{--}} \\
	F_2(\phi) &=& \dfrac{5\sqrt{\pi}}{96}\eta^{*} = \dfrac{5\sqrt{\pi}}{96}\left(\eta_k^{*} + \eta_c^{*} + \eta_b^{*} \right) &  \multicolumn{3}{c}{\text{--}}\\
	F_3(\phi) &=& \dfrac{225\sqrt{\pi}}{1152} \kappa* = \dfrac{225\sqrt{\pi}}{1152} \left(\kappa_k^{*} + \kappa_c^{*} + \kappa_b^{*} \right) &   \multicolumn{3}{c}{\text{--}}\\
	F_4(\phi) &=& \dfrac{12}{\sqrt{\pi}}(1-e^2)\phi^2g_0(\phi) & F_4^{\prime}(\phi) &=& \dfrac{12}{\sqrt{\pi}}(1-e_{eff}^2)\phi^2g_0(\phi)\\
	\eta_k^{*} &=& \dfrac{1-2/5(1+e)(1-3e)\phi g_0(\phi)}{\left(1-1/4(1-e)^2-5/24(1-e^2)\right)g_0(\phi)} &  \eta_k^{*} &=& \dfrac{48/(5\sqrt{\pi})\phi-2/5(1+e)(1-3e)\phi g_0(\phi)}{\left(1-1/4(1-e)^2-5/24(1-e^2)\right)g_0(\phi)} \\
	\eta_c^{*} &=& \dfrac{4}{5}(1+e)\phi g_0(\phi)\eta_k^{*} &  \multicolumn{3}{c}{\text{--}} \\
	\eta_b^{*} &=& \dfrac{384}{25\pi}(1+e)\phi^2g_0(\phi)&  \multicolumn{3}{c}{\text{--}} \\
	\kappa_k^{*} &=& \dfrac{2\left[1+3/5(1+e)^2(2e-1)\phi g_0(\phi)\right]}{(1-7/16(1-e))(1+e)g_0(\phi)}&  \kappa_k^{*} &=& \dfrac{2\left[576/(225\sqrt{\pi})\phi +3/5(1+e)^2(2e-1)\phi g_0(\phi)\right]}{(1-7/16(1-e))(1+e)g_0(\phi)} \\
	\kappa_c^{*} &=& \dfrac{6}{5}(1+e)\phi g_0(\phi)\kappa_k^{*} &  \multicolumn{3}{c}{\text{--}}\\
	\kappa_b^{*} &=& \dfrac{2304}{225\pi}(1+e)\phi^2g_0(\phi) &  \multicolumn{3}{c}{\text{--}} 
	\end{array}$}

	\caption{Expression of the coefficients in the constitutive relations of the kinetic theory given by the \cite{garzo1999} model (left column) and the proposed modifications (right column).}
	\label{tab::kin_th}
\end{table}

In these relations, a key parameter is the radial distribution function $g_0(\phi)$. It characterizes the degree of spatial correlation of two particles that are going to collide. It is therefore expected that $g_0$ is one in the dilute regime (no spatial correlation) and diverges when $\phi$ approaches $\phi_{max}$ (full spatial correlations). The first functional form has been proposed by \citet{carnahan1969}, as $g_0(\phi) = (2-\phi)/(2(1-\phi)^3)$. It gives very good results for low values of packing fraction but does not diverge in the dense limit. To correct this behavior, other functional forms have been proposed theoretically \citep{ma1986, torquato1995} and empirically \citep{lun1986, chialvo2013, vescovi2014}. In the following, the form of \citet{chialvo2013} proposed based on discrete element simulations of simple shear flows, is adopted,
\begin{equation}
g_{0}(\phi) = \dfrac{2-\phi}{2(1-\phi)^3} + \dfrac{a \phi^2}{(\phi_{max} -\phi)^{3/2}},
\label{eq::g0Ch}
\end{equation}
where $a=0.58$ is an empirical coefficient. This expression has the advantage of matching the \citet{carnahan1969} radial distribution function for low packing fraction and to diverge close to $\phi_{max}$.\\

The last term for which a closure is required is $J_{int}$. It corresponds to the work done by the drag force due to fluctuating particle motion. Assuming that the drag is linear in relative velocity (i.e. $K \sim Const$), it can be expressed as \citep{ding1990, fox2014}
\begin{equation}
J_{int} = \phi(1-\phi)K(\left<u_i^{f\prime}v_i^{p\prime}\right>^s-3T).
\label{eq::Jintinit}
\end{equation}
The second term in the parenthesis of equation~\eqref{eq::Jintinit} is a dissipation term. The first term is a source term representing a transfer of fluctuating energy from the fluid to particles through turbulence. It depends on the degree of correlation between fluctuating velocities of both phases and typically varies from $0$ (uncorrelated motions) to $2k$ (fully correlated motions), where $k$ is the fluid fluctuating kinetic energy \citep{danon1977, chen1985}. In the DEM model that is used in this study (see section~\ref{sec::dem}), no fluid velocity fluctuations are computed and there is therefore no correlation between the fluid and granular phase fluctuation velocities. This is equivalent to consider that particles are very inertial and that their motion is not at all influenced by the turbulent structures of the fluid flow. The drag term therefore simplifies as
\begin{equation}
J_{int} = -\phi(1-\phi)K\times3T,
\label{eq::Jintlin}
\end{equation}
and it represents the dissipation of granular temperature by the drag force.

%%%%%%%%%%%%%%%%%%%%%%%%%%%%%%%%%%%%%%%%
%%% DEM Model
%%%%%%%%%%%%%%%%%%%%%%%%%%%%%%%%%%%%%%%%

\subsection{Euler-Lagrange model}
\label{sec::dem}

The continuum two-fluid model will be tested against coupled fluid-DEM simulations. This is a three-dimensional DEM model, based on the open-source code YADE \citep{smilaueretal.2015}, coupled with a one-dimensional turbulent fluid flow model. It is presented in \cite{maurin2015, maurin2015a} and has been validated with experiments \citep{frey2014}. It will be briefly presented herein but the interested reader is referred to \citet{maurin2015} and \citet{maurin2015a} for a complete description. 

Frictional spheres of density $\rho^p$ and diameter $d$ submitted to a gravity acceleration $g = 9.81$ $m.s^{-2}$ are considered. A linear spring-dashpot model \citep{schwager2007} composed of a spring of stiffness $k_n$, computed to stay in the rigid limit of grains \citep{roux2002}, in parallel with a viscous damper of coefficient $c_n$ in the normal direction; and a spring of stiffness $k_s=k_n$ associated with a slider of friction coefficient $\mu^p$ in the tangential direction. The normal stiffness, in parallel with the viscous damper, defines a normal restitution coefficient $e$. In addition, each particle is submitted to a buoyancy and a fluid drag force. The same drag model as for the two-fluid model is considered (eq.\eqref{eq::dragForce} and \eqref{eq::dragCoef}).

For the fluid phase, a one-dimensional (1D) vertical model is used, in which the fluid velocity is only a function of the wall-normal component, $z$, and is aligned with the streamwise direction. This is the exact same fluid model as presented previously for the two phase flow model (eq.~\eqref{eq::momxFluid}). The only difference lies in the drag interaction term. Indeed as a fluid drag force applies to each individual particle, it is necessary to compute the average momentum transferred by the fluid to the particles through the drag force, 
\begin{equation}
n\left< \bm{f_D} \right>^s = \dfrac{3\rho^f}{4d}\phi\left(1-\phi\right)^{-\zeta} \left< C_D ||\bm{u}^f - \bm{u}^p||(\bm{u}^f - \bm{u}^p)\right>^s,
\end{equation}
where $\left< \cdot \right>^s$ denotes the solid phase average. The averaging procedure proposed by \citet{jackson1997,jackson2000} is adopted. It is based on a weighting function, representing the volume in which the averaging is performed. Considering the symmetry of the problem, a cuboid weighting function $\mathcal{H}$ with the same length and width as the 3D domain is applied. In the vertical direction, in order to capture the strong gradient of the mean flow, the vertical thickness of the box is taken as $d_z = d/30$. The volume fraction is computed as
\begin{equation}
\phi(z) = \sum_p \int_{V_p}\mathcal{H}(|z-z^{\prime}|)dV,
\end{equation}
where $V_p$ represents the volume of particle $p$. The average over the solid phase of any scalar quantity $\gamma$ is computed as
\begin{equation}
\left<\gamma\right>^s(z) = \dfrac{1}{\phi(z)}\sum_p \int_{V_p}\gamma(z^{\prime})\mathcal{H}(|z-z^{\prime}|)dV.
\end{equation}
This averaging procedure makes it possible to compute the total transfer of momentum between fluid and particles through the drag force and to obtain vertical profile of all particle quantity (i.e. particle velocity profile).

\citet{goldhirsch2010} and later \citet{pahtz2015} proposed a rigorous procedure to compute the DEM granular stresses $\sigma^p$, granular temperature flux $q$, dissipation during collision $\Gamma$ and drag dissipation $J_{int}$. For the same weighting function as previously, they are expressed as
\begin{align}
\label{eq::sigmaDEM}
\sigma^p_{ij}(z) &= -\rho^p\phi\left<v_i^{p\prime}v_j^{p\prime} \right>^s - \dfrac{1}{V}\sum_c f_i^cb_j^c,\\
\label{eq::diffDEM}
q(z) &= - \dfrac{1}{2}\rho^p\phi\left<v_z^{p\prime}v_i^{p\prime}v_i^{p\prime}\right>^s - \dfrac{1}{V}\sum_c f_i^cv_i^pb_z^c,\\
\label{eq::gammaDEM}
\Gamma(z) &= -\dfrac{1}{2V}\sum_{p, q|z^p\in[z-d_z/2, z+d_z/2]}f_i^c(v_i^p -v_i^q),\\
\label{eq::JDEM}
J_{int} &= n\left<f_{Di}v_i^{p\prime}\right>^s,
\end{align}
where the sums are performed over the ensemble of particles $p$, particles $q$ in contact with $p$ and over contacts $c$. $\bm{f^c}$ is the interaction force at contact $c$ applied on particle $p$ by particle $q$ and $\bm{b}^c = \bm{x}^q-\bm{x}^p$ is the branch vector. In these expressions, the Einstein summation has been used. Note that the granular stress tensor and granular temperature flux are the sum of kinetic (similar to a Reynolds stress for a fluid) and contact contributions.\\

%%%%%%%%%%%%%%%%%%%%%%%%%%%%%
%%%% Results %%%%%%%%%%%%%%%%
%%%%%%%%%%%%%%%%%%%%%%%%%%%%%
\section{Results}\label{sec::results}

In this section, coupled fluid-discrete simulations will be used to compare with two-fluid simulations, showing important departures between both models. The origins of the discrepancies will be physically investigated and corrections to the KT model will be proposed in order to reproduce quantitatively the DEM simulations.

\subsection{Euler-Lagrange simulation results}\label{sec::discrete_sim}

\begin{figure}
	\centering
	\subfloat[]{\includegraphics[height=3.8cm]{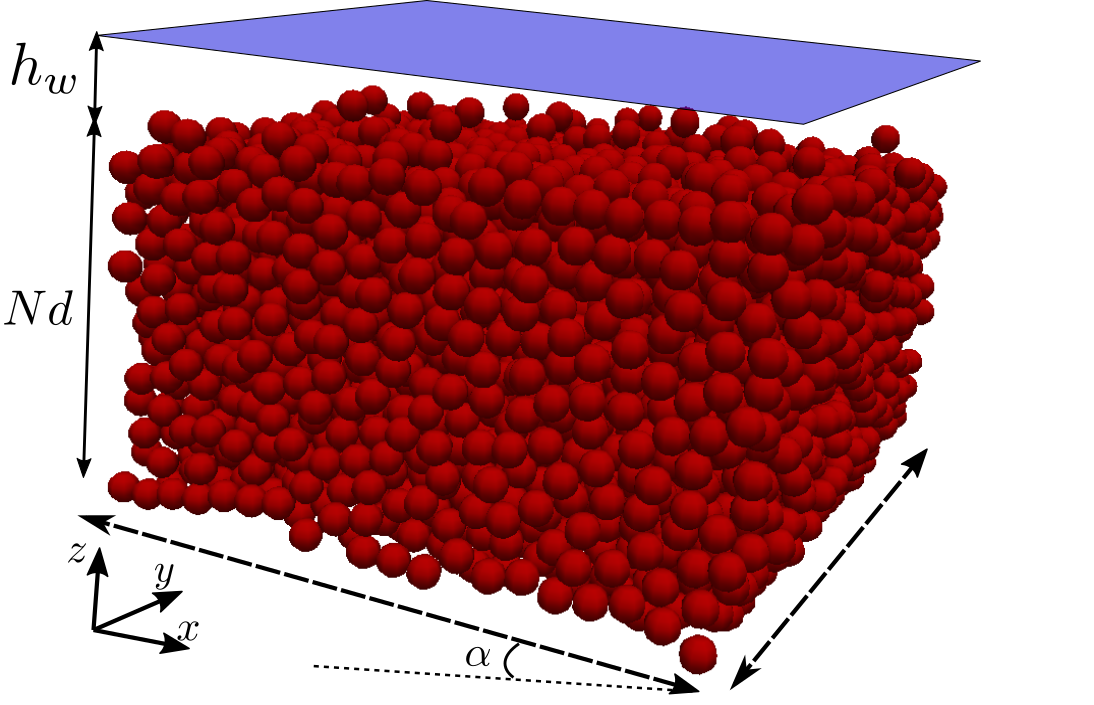}}
	\subfloat[]{\includegraphics[height=3.8cm]{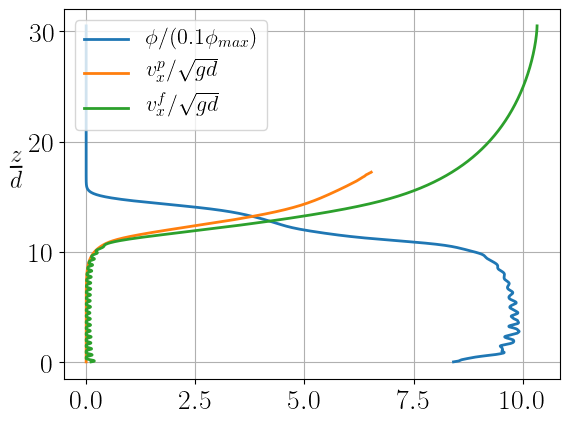}}
	\subfloat[]{\includegraphics[height=3.8cm]{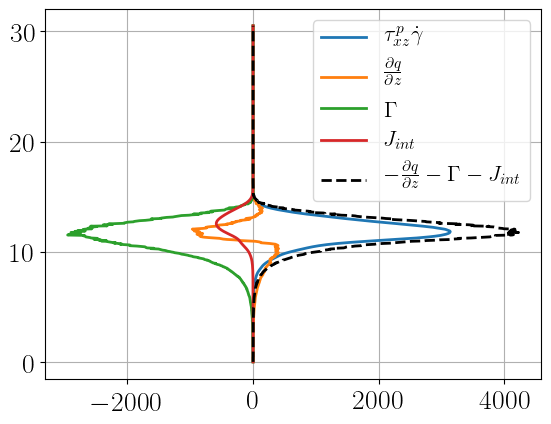}}
	\caption{Fluid-DEM simulation example for $\theta = 0.6$, $e=0.7$, $\mu^p=0.4$. (a) Numerical setup of the DEM simulations. $N=12$ layers of particles are deposited by gravity above a rough bed made of fixed particles. The fluid free surface is set to $H_f = Nd+h_w$. (b) Packing fraction profile normalized for visibility by one tenth of $\phi_{max} = 0.635$ and dimensionless particle and fluid velocity profiles. (c) Granular temperature balance in Pa.s\textsuperscript{-1}.}
	\label{fig::setup}
\end{figure}

The numerical setup is presented in figure~\ref{fig::setup}a. Particles of diameter $d=6$~mm and density $\rho^p=2500$ kg.m\textsuperscript{-3} are initially deposited by gravity over a rough bed made of fixed particles. Particles are immersed in a fluid of density $\rho^f=1000$ kg.m\textsuperscript{-3}. The free surface is fixed to $H_f = H_{bed} + h_w$, where $H_{bed}=12.5d$ represents the bed height at rest and $h_w$ represents the water depth. The slope is fixed to $\sin(\alpha) = 5\%$ ($2.85^\circ$).  This defines the Shields number as $\theta = \rho^fgh_w\sin(\alpha)/\left[(\rho^p-\rho^f)gd\right]$. At initial time, the fluid flows by gravity and sets particles into motion. After a short transient, during which fluid and particles are accelerating, a steady state takes place at transport equilibrium. A set of simulations have been performed in which the Shields number is varied from $\theta=0.2$, close to the threshold of motion, until $\theta=1$ corresponding to intense bedload transport. Three restitution coefficients $e=0.5$, $e=0.7$ and $e=0.9$ have been explored and frictional particles are considered with $\mu^p=0.4$. The parameters of all simulations are summarized in table~\ref{tab::param}.

Figure~\ref{fig::setup}b shows typical profiles of packing fraction, particle and fluid velocity for the case $\theta=0.6$, $e=0.7$. The sediment phase shows a very complex phenomenology, with continuous transition from a pure fluid phase ($z>17d$) to a quasi-static creeping regime ($z<8d$). This transition zone is called the bedload layer in which most of the transport is taking place. At the same time, both the fluid and particle velocity decrease when going down into the bed. The quasi-static regime, below $z<8d$, is characterized by a roughly constant maximal packing fraction $\phi = \phi_{max}$ and very slow fluid and granular motions with exponentially decreasing velocity profiles with depth \citep{houssais2015, chassagne2020a}. The quasi-static regime has a negligible contribution to the overall transport and will be neglected in this work.

Figure~\ref{fig::setup}c shows the profiles of granular temperature budget (eq.~\eqref{eq::TPart}): production, diffusion, dissipation through collisions and drag dissipation term. At steady state, the left hand side of the granular temperature equation~\eqref{eq::TPart} vanishes and the production of temperature should be balanced by diffusion, contact dissipation and drag dissipation. As it can be observed in figure~\ref{fig::setup}c, the black dashed line representing the sum of these three terms is not superimposed to the production term indicating that the balance is not completely closed. \citet{pahtz2015} had the same issue in their DEM simulations and attributed it to the dissipation term $\Gamma$ due to the lack of scale separation in the averaging procedure, i.e. the vertical discretization $d_z = d/30 << d$. For this reason in the following, the dissipation term will be computed in order to close the granular temperature balance equation, $\Gamma = -\tau_{xz}^p\dot{\gamma} - \partial q/\partial z - J_{int}$. 

\begin{table}
	%\resizebox{\textwidth}{!}{
	$
		\begin{array}{|c|c|c|c|c|}
		%&\multicolumn{3}{c|}{\text{DEM parameters}} & \multicolumn{3}{c}{\text{Two-fluid simulation parameters}}\\
		\hline
		\theta & h_w & H_f & N_{cell} & \Delta z \\
		\hline
		0.2 & 6d & 18.5d & 200 & 0.0925d\\
		0.4 & 12d & 24.5d & 120 & 0.204d\\
		0.6 & 18d & 30.5d & 120 & 0.254d\\
		0.8 & 24d & 36.5d & 200 & 0.183d\\
		1 & 30d & 42.5d & 200 & 0.213d\\
		\hline
		\end{array}
		$
		%$}
	
	\caption{Physical and numerical parameters for each configuration. Shields number, fluid depth, free surface position, number of cells and vertical discretization in sedFoam simulations.}
	\label{tab::param}
\end{table}

\subsection{Direct comparison with the two-fluid simulations}

\begin{figure}
	\centering
	\includegraphics[width=\linewidth]{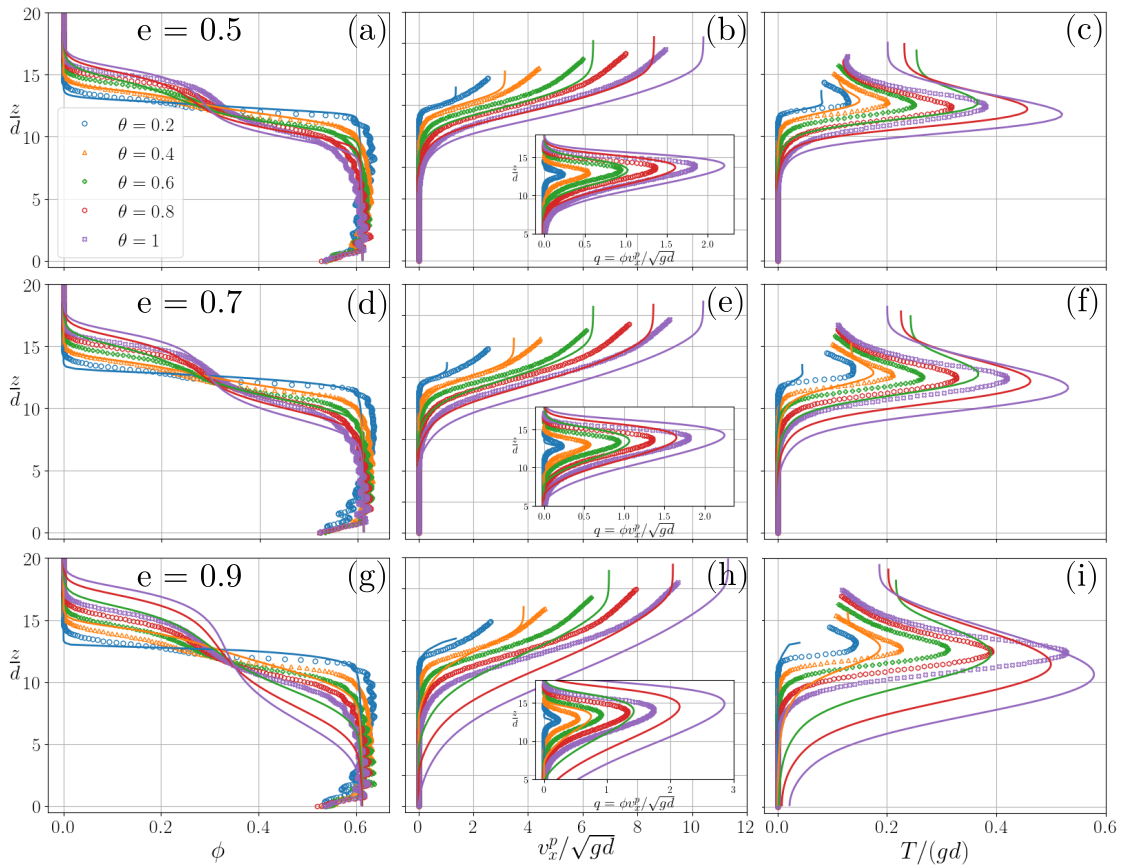}
	%\subfloat[]{\includegraphics[height=4.1cm]{phiAlle05}}
	%\subfloat[]{\includegraphics[height=4cm]{vxAlle05}}
	%\subfloat[]{\includegraphics[height=4cm]{TAlle05}}\\
	%\subfloat[]{\includegraphics[height=4.1cm]{phiAlle07}}
	%\subfloat[]{\includegraphics[height=4cm]{vxAlle07}}
	%\subfloat[]{\includegraphics[height=4cm]{TAlle07}}\\
	%\subfloat[]{\includegraphics[height=4.1cm]{phiAlle09}}
	%\subfloat[]{\includegraphics[height=4cm]{vxAlle09}}
	%\subfloat[]{\includegraphics[height=4cm]{TAlle09}}
	\caption{Comparison between DEM (symbols) and two phase flow results with initial model (full line) for $e=0.5$ (first row), $e=0.7$ (second row) and $e=0.9$ (third row). (a-d-g) Packing fraction profiles, (b-e-h) particle velocity and transport (inset) profiles and (c-f-i) granular temperature.}
	\label{fig::resInit}
\end{figure}

The two-fluid model can now be compared with the results of the coupled fluid-DEM model. The fluid equations being the same in both models, a particular focus will be given to the granular behavior of the particle bed. The same setup as presented in section~\ref{sec::discrete_sim} is simulated using the two-fluid model, with parameters reported in table~\ref{tab::param}. The mesh is composed of $120$ or $200$ cells depending on the configuration (see table~\ref{tab::param}) in order to maintain a grid resolution between $0.1d$ and $0.25d$. The numerical implementation is fully described in \citet{chauchat2017}, consisting of a Total Variation Diminution (TVD) central scheme, with \textit{Euler implicit} first order scheme for time derivatives, \textit{Gauss linear} scheme for spatial gradient operators, \textit{Gauss limitedLinear} for divergence operators and \textit{Gauss linear corrected} for laplacian operators. The details of these schemes can be found in the OpenFOAM documentation and they are all second order in space. The boundary conditions are set to \textit{cyclic} in the stream-wise direction and \textit{empty} in the vertical direction. At the top boundary, the pressure of both phases is fixed to $0$ and a zero gradient is imposed to all other fields. The bottom boundary condition is set to \textit{fixedFluxPressure} for the pressure and to \textit{noslip} for the velocities. The CFL is fixed to $0.1$ with a maximal time step $\Delta t_{max} = 10^{-3}$s. \\

Figure~\ref{fig::resInit} compares, for all Shields numbers and for the three restitution coefficients, the profiles of packing fraction, particle velocity and granular temperature computed with the DEM (symbols) and with the two-fluid model (full lines). The two-fluid model mispredicts the shape of the packing fraction profile and overestimates strongly the particle velocity and the granular temperature for all Shields numbers. The observed errors are more important for larger restitution coefficients. The bed flows deeper in the two-fluid simulations than in the DEM results and a plateau of velocity in the dilute part of the granular flow is predicted which is not observed in the DEM results. As a consequence the width of the bedload layer is larger in the continuum model than in the DEM (see inset of figure~\ref{fig::resInit}b). Similarly, the granular temperature is overestimated  by the continuum model, in particular inside the granular bed, suggesting that granular fluctuating energy dissipation is underestimated in the continuum simulations. As discussed in the introduction, this is because particles are frictional and dissipate more energy than what is predicted in the frictionless KT model.

In order to improve the results of the two-fluid model, some corrections to the \citet{garzo1999} model are needed. In particular this model does not account for interparticle friction while, as it will be shown, it plays an important role in the dynamics of the granular behavior. In addition, the departure between both models in the dilute part of the granular flow indicates that saltation plays an important role in the dynamics of the flow. Following \citet{chialvo2013}, the DEM results will be used as a guideline to determine modified coefficients of the KT to account for frictional interactions and saltation.

\subsection{Corrections to the \citet{garzo1999} model}
\label{sec::newmod}

Following equations~\eqref{eq::sigmaDEM}, \eqref{eq::diffDEM}, \eqref{eq::gammaDEM} and \eqref{eq::JDEM}, all terms of the granular temperature equation can be computed in the DEM and the KT laws (eq~\eqref{eq::pkin}, \eqref{eq::etakin}, \eqref{eq::kappakin}, \eqref{eq::gammakin} and \eqref{eq::Jintlin}) can be compared with the DEM results. Additionally to previous simulations, a DEM simulation with frictionless particles ($\mu^p=0$, $\theta=0.4$, $e=0.7$) have also been performed to investigate the role of friction on the granular behavior. \\

\begin{figure}
	\centering
	\subfloat[]{\includegraphics[height=6.3cm]{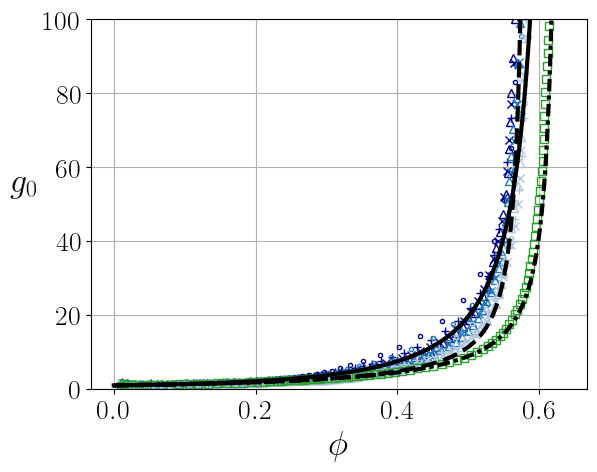}}
	\subfloat[]{\includegraphics[height=6.3cm]{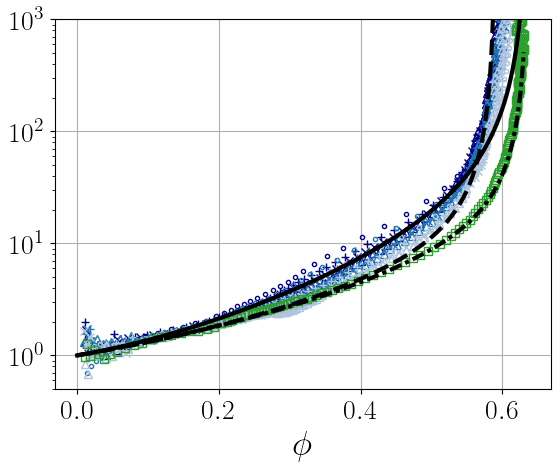}}
	\caption{Radial distribution function computed in the DEM simulations : $\theta=0.4$ ($\circ$), $\theta=0.6$ ($+$), $\theta=0.8$ ($\times$) and $\theta=1$ ($\bigtriangleup$) for three restitution coefficients $e=0.5$ (dark blue), $e=0.7$ (medium blue) and $e=0.9$ (light blue). Green square corresponds to frictionless particles $\theta=0.4$ and $e=0.7$. Lines correspond to equation~\eqref{eq::g0Ch} for :
		$a=0.58$, $\phi_{max}=0.635$ (\protect\tikz[baseline]{\protect\draw[line width=0.3mm,densely dash dot] (0,.5ex)--++(0.52,0) ;});
		$a=0.58$, $\phi_{max}=0.59$ (\protect\tikz[baseline]{\protect\draw[line width=0.3mm,densely dashed] (0,.5ex)--++(0.45,0) ;});
		$a=2.71$, $\phi_{max}=0.635$ (\protect\tikz[baseline]{\protect\draw[line width=0.3mm] (0,.5ex)--++(0.5,0) ;})}.
	\label{fig::g0}
\end{figure}

As mentioned previously, the radial distribution function is a key parameter of the KT. From the DEM simulations, it can be estimated using the expression of $F_1$ and by reversing the pressure law (eq.~\eqref{eq::pkin}) as,
\begin{equation}
g_0(\phi) = \dfrac{1}{2\phi(1+e)}\left(\dfrac{p^{kin}}{\rho^p\phi T}-1\right).
\label{eq::g0}
\end{equation}
In the DEM simulations, it is not possible to isolate the kinetic contribution from the elastic one in the granular pressure. The radial distribution function is therefore computed in equation~\eqref{eq::g0} based on the total pressure. The elastic pressure is however predominant only for packing fraction very close to the random close packing $\phi_{max} \sim 0.635$ and it is assumed that assimilating the kinetic pressure to the total pressure has a negligible effect on the results. Figure~\ref{fig::g0} shows the computed radial distribution function in the DEM simulations (symbols) compared with classical expressions of the literature (black lines). The DEM data show two distinct trends for frictional and frictionless particles. The radial distribution function for frictionless particles is in perfect agreement with the form proposed by \citet{chialvo2013} (eq.~\eqref{eq::g0Ch}, dashed dotted line). This is quite remarkable because expression~\eqref{eq::g0Ch} were obtained in very different configuration -- dry granular flow in simple shear cell -- showing that our DEM simulations are in line with the literature on this point. \citet{vescovi2014} proposed another expression of the radial distribution function for frictionless spheres that is almost superimposed with equation~\eqref{eq::g0Ch} (not shown in figure~\ref{fig::g0} for clarity).

The DEM data show that inter-particle friction impacts the radial distribution function. As observed by \citet{dacruz2005}, \citet{chialvo2012} and \citet{chialvo2013} in shear cell numerical simulations, the radial distribution function diverges at lower values of packing fraction. To account for this effect, \citet{chialvo2013} simply suggests to let the critical packing fraction depend on interparticle friction in equation~\eqref{eq::g0Ch} with $\phi_{max}(\mu^p=0.4) = 0.59$. Figure~\ref{fig::g0} (dashed line) however shows that it is not sufficient to capture the increase of $g_0$ for intermediate values of $\phi$. In addition, in the bedload configuration, the packing fraction will always exceed $0.59$, even for frictional particles (see figure~\ref{fig::setup}b). This is because in the bed, there is a transition from kinetic (collisional) stresses to elastic stresses. Fitting the DEM data, and based on the form of \citet{chialvo2013}, it is proposed to modify the empirical coefficient in equation~\eqref{eq::g0Ch} to $a=2.71$ and letting $\phi_{max}=0.635$. This is plotted in figure~\ref{fig::g0} (full line) showing good agreements with the DEM data. \\

\begin{figure}[h!]
	\centering
	\subfloat[]{\includegraphics[height=6.cm]{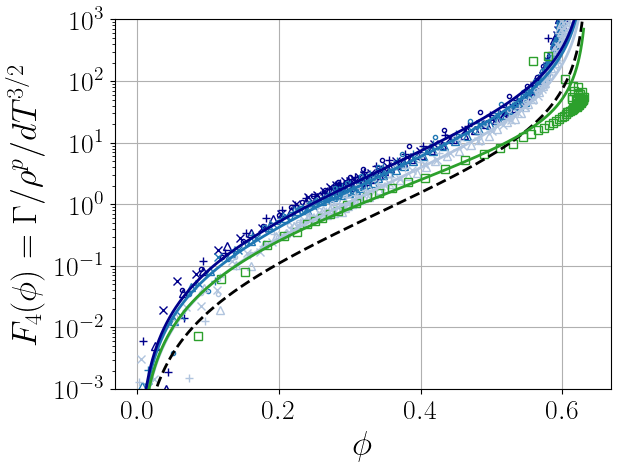}}
	\subfloat[]{\includegraphics[height=6.cm]{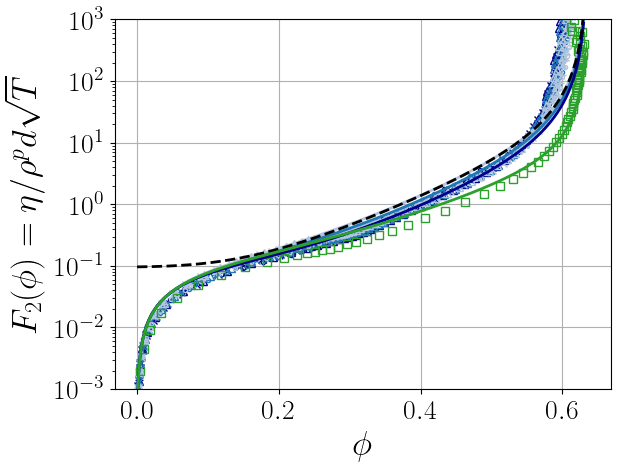}}\\
	\subfloat[]{\includegraphics[height=6.cm]{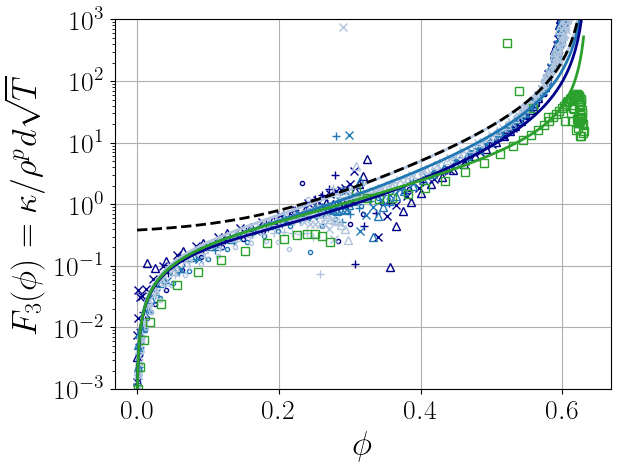}}
	\subfloat[]{\includegraphics[height=6.cm]{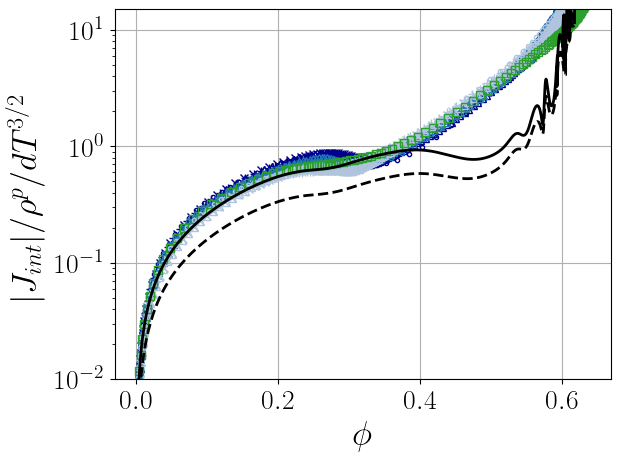}}
	\caption{KT coefficients computed in the DEM (symbols, same as in figure~\ref{fig::g0}) 
		compared with the modified KT coefficients (colored full line) for frictionless (green) and frictional case with $e=0.9$ (dark blue), $e=0.7$ (medium blue) and $e=0.5$ (light blue). Laws with initial model and $e=0.9$ are plotted for comparison (\protect\tikz[baseline]{\protect\draw[line width=0.3mm,densely dashed] (0,.5ex)--++(0.45,0) ;}). 
	 (a) dimensionless collisional dissipation, (b) dimensionless viscosity, (c) dimensionless heat transport coefficient. (d) Dimensionless granular temperature dissipation by the drag force compared with equation~\eqref{eq::Jintlin} (\protect\tikz[baseline]{\protect\draw[line width=0.3mm,densely dashed] (0,.5ex)--++(0.45,0) ;}) and with corrected model (eq.~\ref{eq::Ttransfer}) (\protect\tikz[baseline]{\protect\draw[line width=0.3mm] (0,.5ex)--++(0.5,0) ;}).}
	\label{fig::kin_th_plot}
\end{figure}

With the obtained radial distribution functions for frictionless ($a=0.58$) and frictional ($a=2.71$) particles, it is possible to compare the KT closure laws listed in the table~\ref{tab::kin_th} with the DEM results. This is done in figure~\ref{fig::kin_th_plot}, where coefficients $F_2$, $F_3$, $F_4$ and $J_{int}$ are plotted as a function of $\phi$. For all coefficients, the data points are superimposed at low packing fraction but separated into two branches at high packing fraction whether particles are frictional or not. This separation is mainly due to the difference of radial distribution function. It indicates that interparticle friction has a stronger effect than the restitution coefficient on the KT closure laws. \\
%For visibility, only three curves are plotted in the figures : KT laws for $e=0.9$ and frictional particles with initial \cite{garzo1999} model (dashed line) and after modification (dashed line). The dashed-dotted line is the law for frictionless particles ($a=0.58$ in $g_0$).\\

The previous discussion has shown that interparticle friction impacts strongly modifies the radial distribution function. It is also expected to influence granular temperature dissipation. Indeed, at contact between particles, energy is dissipated through inelasticity and friction. The \citet{garzo1999} model was derived for frictionless spheres and therefore does not account for this second source of dissipation. This is clearly evidenced in figure~\ref{fig::kin_th_plot}a showing the dimensionless dissipation computed in the DEM simulations. Dissipation is clearly higher for frictional (blue symbols) than for frictionless particles (green squares). Dissipation is a bit higher for lower restitution coefficient (compare dark blue to light blue symbols) but it is more sensitive to interparticle friction. The dissipation for frictionless particles is very well predicted by the \citet{garzo1999} model (green line) but not for frictional particles (dashed line underestimates light blue symbols). To account for inter-particle friction, \citet{jenkins2002} proposed to treat it similarly to inelasticity by a modification of the restitution coefficient $e_{eff}(\mu^p) \leq e$ where $e_{eff}$ is the effective restitution coefficient. Based on discrete simulations, \citet{chialvo2013} proposed the following empirical expression,
\begin{equation}
e_{eff} = e - f(\mu^p) = e - \dfrac{3}{2}\mu^p \exp{(-3\mu^p)}.
\end{equation}
The modified coefficient $F_4^{\prime} = F_4 (1-e_{eff}^2)/(1-e^2)$ is plotted in full colored lines for the three restitution coefficient and is now in very good agreements with the DEM results.\\
%The $F_1$ function perfectly predicts the dimensionless pressure in the DEM simulations (see figure~\ref{fig::kin_th_plot}a). This is not surprising as we fitted the $g_0$ function using the expression of $F_1$. 

The dimensionless viscosity and heat transport coefficients are plotted in figures~\ref{fig::kin_th_plot}b and c. The \citet{garzo1999} coefficients are plotted in black dashed lines for frictional particles and $e=0.9$. They are quantitatively good for large packing fraction but fails at reproducing the DEM data at low packing fraction. In this limit, the kinetic contributions $\eta_k^*$ and $\kappa_k^*$ of both coefficients are predominant and the observed discrepancies therefore originate from these contributions. Indeed, both $\eta_k^*$ and $\kappa_k^*$ contain a non dependent term with $\phi$ (see table~\ref{tab::kin_th}) and the dimensionless viscosity and heat transfer coefficients are non-vanishing in the very dilute limit contrary to what is observed in the DEM simulations. This discrepancies between theory and the DEM coefficients is at the origin of the difference of behavior of particle velocity in the dilute regime (see figure~\ref{fig::resInit}b). In particular, the constant viscosity for $\phi$ close to zero leads to the formation of the velocity plateau. 

We attribute this departure from the \citet{garzo1999} model to saltation which is the main mode of transport in the dilute part of the flow. Particles have ballistic trajectories that are controlled by gravity and drag force and therefore can not be described by the KT. \citet{jenkins2018} proposed a continuum theory for saltation in aeolian transport in which the granular viscosity is linear with the packing fraction and therefore goes to zero in the dilute regime, as observed in our DEM simulations. In the DEM data, it is possible to isolate the kinetic contributions (first term in the right hand side of equations~\eqref{eq::sigmaDEM} and \eqref{eq::diffDEM}) and, following \citet{jenkins2018}, the constant parts of $\eta_k^*$ and $\kappa_k^*$ are made linear with $\phi$ as 
\begin{align}
\eta_k^{*} &= \dfrac{48/(5\sqrt{\pi})\phi-2/5(1+e)(1-3e)\phi g_0(\phi)}{\left(1-1/4(1-e)^2-5/24(1-e^2)\right)g_0(\phi)},\\
\kappa_k^{*} &= \dfrac{2\left[576/(225\sqrt{\pi})\phi +3/5(1+e)^2(2e-1)\phi g_0(\phi)\right]}{(1-7/16(1-e))(1+e)g_0(\phi)}.
\end{align}
Note that these modifications also impact the contact contributions $\eta_c^{*}$ and $\kappa_c^{*}$ as they are proportional to the kinetic contributions. The modified coefficients are plotted in figures~\ref{fig::kin_th_plot}b,c for frictional (blue lines) and frictionless (green line) particles and they perfectly reproduce the DEM data. The proposed empirical modifications should be verified in other configurations and parameterized following physical arguments. This would represent a complete study that lie outside of the present work objectives. However, it suggests that the modelling of saltation, \textit{a priori} out of the scope of the KT, could be included in this framework.\\

The work done by the drag force due to granular fluctuating motion $J_{int} = n\left<f_{Di}v_i^{p\prime}\right>^s$ is plotted in figure~\ref{fig::kin_th_plot}d. As expected, no dependency on the restitution coefficient nor on interparticle friction is observed. The black dashed line corresponds to the closure of \citet{ding1990} in equation~\eqref{eq::Jintlin}. This closure underestimates the effective dissipation of fluctuating energy by the drag force for the whole range of packing fraction. This is because it was assumed a linear drag force to derive closure~\eqref{eq::Jintlin}, while it is not the case in this configuration (or in any turbulent flow application). To understand how the quadratic nature of the drag force influences the dissipation of fluctuating energy, let us replace by the drag force expression~\eqref{eq::dragForce} in $J_{int}$,

\begin{equation}
J_{int} = n\left<f_{Di}v_i^{p\prime}\right>^s = \dfrac{3}{4}\dfrac{\phi\rho^f(1-\phi)^{-\zeta}}{d}\left<C_D||\bm{u^f} - \bm{v^p}||(u_i^f - v_i^p)v_i^{p\prime}\right>^s.
\end{equation}
\begin{equation}
J_{int} \propto \left<C_D||\bm{u^f} - \bm{v^p}||(u_i^f - v_i^p)v_i^{p\prime}\right>^s.
\end{equation}
In order to provide a closure, it is necessary to discuss the term inside the averaging operator $\left< \right>^s$. Indeed, depending on the value of the particle Reynolds number $Re_p$, $C_D||\bm{u^f} - \bm{v^p}||(u_i^f - v_i^p)$ can be linear with the relative velocity (laminar regime), quadratic (turbulent regime) or in between. While it is usually assumed that the drag is linear \citep{ding1990, fox2014}, this assumption can not be done in the bedload transport configuration. Indeed, the particle Reynolds number ranges between $Re_p \sim 100$ in the bed to $Re_p \sim 2500$ at the bed surface. The drag force is therefore quadratic at the surface and in a transitional regime in the bed. Splitting the particle velocity into a mean and fluctuating component and performing a Taylor expansion to first order with respect to the mean values, the following expression of the granular temperature dissipation through the drag force is obtained (see appendix):
\begin{equation}
J_{int} \sim - \phi(1-\phi)K\left(3+2\dfrac{C_D^{\infty}}{C_D}\right)T,
\label{eq::Ttransfer}
\end{equation}
where $C_D^{\infty}$ is the value of the drag coefficient for an infinite particle Reynolds number. Note that this term is negative and corresponds to a loss of energy for the granular phase. The term $2C_D^{\infty}/C_D$ represents an additional source of dissipation due to the quadratic nature of the drag force. In order to derive this expression it has been necessary to assume that $|| \bm{u^f} - \left<\bm{v^p}\right>^s|| \sim | u_x^f - \left< v_x^p\right>^s|$ and that $\left<v_x^{p\prime}v_x^{p\prime}\right>^s \sim 2T$. These assumptions are specific to configurations with a preferential shearing direction with anisotropic velocity fluctuations. However computations presented in appendix can be easily adapted to other cases. The corrected closure~\eqref{eq::Ttransfer} is plotted in colored full lines and compared with DEM data in figure~\ref{fig::kin_th_plot}. It improves quantitatively the prediction of dissipated energy. It shows a very good agreement with the DEM in the dilute regime but underestimates slightly at high packing fractions. In this limit, the influence of the drag is however very small and this underestimation should not affect the balance of granular temperature.\\

All corrections to the \citet{garzo1999} model proposed in this section are summarized in the right column of table~\ref{tab::kin_th}. The corrected model can now be used to compare with the DEM simulations.

\subsection{Evaluation of the corrected KT model} 

\begin{figure}[h!]
	\centering
	\includegraphics[width=\linewidth]{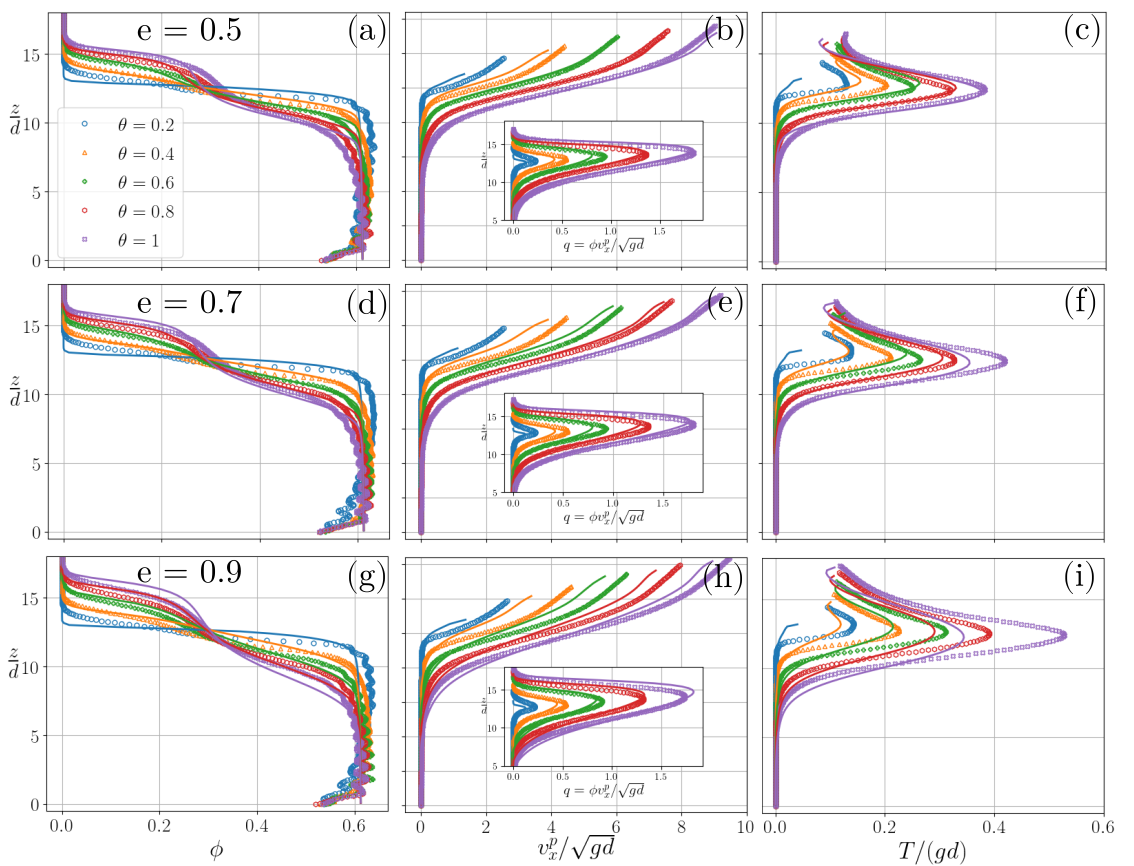}
	%\subfloat[]{\includegraphics[height=4cm]{phiAlle05Mod}}
	%\subfloat[]{\includegraphics[height=4cm]{vxAlle05Mod}}
	%\subfloat[]{\includegraphics[height=4cm]{TAlle05Mod}}\\
	%\subfloat[]{\includegraphics[height=4cm]{phiAlle07Mod}}
	%\subfloat[]{\includegraphics[height=4cm]{vxAlle07Mod}}
	%\subfloat[]{\includegraphics[height=4cm]{TAlle07Mod}}\\
	%\subfloat[]{\includegraphics[height=4cm]{phiAlle09Mod}}
	%\subfloat[]{\includegraphics[height=4cm]{vxAlle09Mod}}
	%\subfloat[]{\includegraphics[height=4cm]{TAlle09Mod}}
	\caption{Comparison between DEM (symbols) and two phase flow results with corrected model (full lines) for $e=0.5$ (first row), $e=0.7$ (second row) and $e=0.9$ (third row). (a-d-g) Packing fraction profiles, (b-e-h) particle velocity and transport (inset) profiles and (c-f-i) granular temperature.}
	\label{fig::resCor}
\end{figure}

The two-fluid model with corrections presented in the previous section is tested against the DEM simulations and the results are presented in figure~\ref{fig::resCor}. Both the packing fraction and velocity profiles are quantitatively improved compared with the initial model (see figure~\ref{fig::resInit}). The results are better for lower restitution coefficients and very acceptable even for $e=0.9$ considering the complexity of the problem. 

The shape of the packing fraction profiles is nicely reproduced, in particular the apparition of the "shoulder" of packing fraction around $\phi \sim 0.3$ for large Shields numbers. The velocity profiles are quantitatively predicted by the continuum model for all Shields numbers and restitution coefficients. In particular, the depth of the flowing layer is perfectly captured. This is quite remarkable as the depth at which the flow stops is a response of the stress and energy balance between both phases and not due to a bottom boundary condition. The plateau of velocity observed previously in the dilute regime (figure~\ref{fig::resInit}) is not present anymore thanks to the correction applied on the viscosity and diffusion laws. As a consequence of the well predicted packing fraction and velocity profiles, the transport profiles are very well reproduced (see inset of figure~\ref{fig::resCor}b and e). The thickness of the transport layer, and its increase with the Shields number, are particularly well predicted. The modified continuum model improves quantitatively the predicted granular temperature (figure~\ref{fig::resCor}c, f and i). It is perfectly predicted for $e=0.5$ but underestimated for larger restitution coefficient. Nevertheless, the results for $e=0.9$ are still acceptable.

%%%%%%%%%%%%%%%%%%%%%%%%%%%%%%%%%%%%%%%%%%%%%
%%% Discussion
%%%%%%%%%%%%%% %%%%%%%%%%%%%%%%%%%%%%%%%%%%%
\section{Discussions} \label{sec::discussion}

The corrected model implemented in SedFoam has been shown to reproduce very accurately the DEM results. The proposed corrections are essentially empirical but the improvements they provide can be analyzed in order to interpret the physics of the granular flow.\\

First, inter-particular friction has been shown to play an important role in the granular behavior. It modifies the radial distribution function and increases dissipation at contact. In our DEM simulations with frictional particles, almost no influence of the restitution coefficient is observed. The packing fraction and velocity profiles are completely superimposed whatever the value of $e$ ranging from $e=0.5$ to $e=0.9$ (compare figure~\ref{fig::resCor}a, d and g or figure~\ref{fig::resCor}b, e and h). It indicates that the dissipation of granular temperature is governed by friction rather than inelasticity of collisions. This is what is observed in figure~\ref{fig::kin_th_plot} where the KT coefficients computed with the DEM are more affected by interparticle friction rather than the restitution coefficient. It is interesting to note that the \citet{garzo1999} model is able to reproduce this property of the flow, with KT coefficients that are more affected by the change of $g_0$ for frictional particles than by the restitution coefficient. This is particularly true for the viscosity and granular temperature diffusivity laws (superimposed full blue lines of figure~\ref{fig::kin_th_plot}b and c), but less evident for the dissipation law. Indeed, $\Gamma$ is proportional to $1-e^2$ and therefore highly dependent on the restitution coefficient when it is large $e>0.7$. When applying the correction proposed by \citet{jenkins2002} and \citet{chialvo2013} which consists in the reduction of the restitution coefficient to account for friction ($e_{eff}<0.7$), the dissipation law is made less dependent on $e$ as illustrated in figure~\ref{fig::kin_th_plot}a (full lines very close for frictional particles). The KT model is therefore applied in a range of parameters that barely depends on the restitution coefficient and is more affected by $g_0$. This highlights the role played by inter-particular friction in the sediment bed dynamics and shows the capacity of the KT model to represent such granular flow.\\

\begin{figure}
	\centering
	\subfloat[]{\includegraphics[height=6cm]{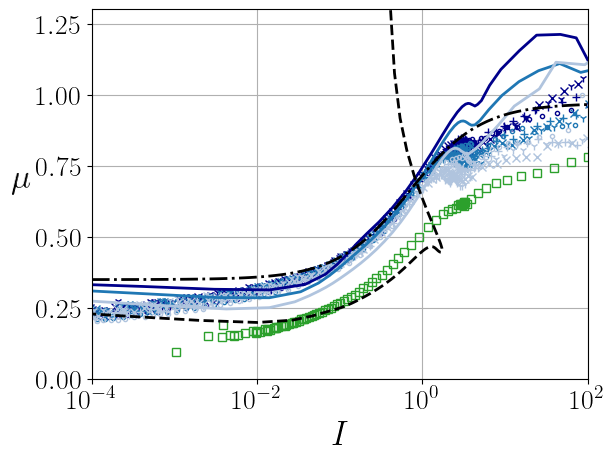}}
	\subfloat[]{\includegraphics[height=6cm]{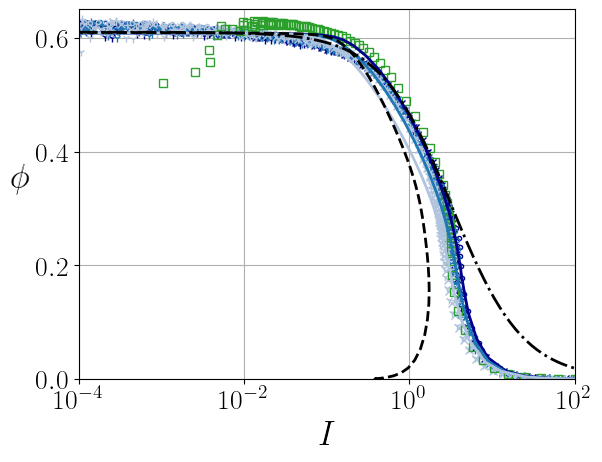}}
	\caption{(a) $\mu(I)$ and (b) $\phi(I)$ relations computed in the DEM (symbols, same as in figure~\ref{fig::g0}) and compared with: 
		the rheological law proposed by \citet{maurin2016} (equations~\eqref{eq::muI} and \eqref{eq::phiI}) for bedload transport (\protect\tikz[baseline]{\protect\draw[line width=0.3mm,densely dash dot] (0,.5ex)--++(0.52,0) ;}) ;
		two phase-flow simulation with the initial model before any correction for the case $e=0.9$, $\mu^p=0.4$ (\protect\tikz[baseline]{\protect\draw[line width=0.3mm,densely dashed] (0,.5ex)--++(0.45,0) ;});
		two phase-flow simulations with the corrected model for the three restitution coefficients and $\theta=0.6$ (\protect\tikz[baseline]{\protect\draw[line width=0.3mm] (0,.5ex)--++(0.5,0) ;}).}
		 %The dashed line corresponds to the case $e=0.9$, $\mu^p=0.4$ with the initial model before any correction. The dotted line indicates the rheological law proposed by \cite{maurin2016} for bedload transport.}
	\label{fig::rheol}
\end{figure}

Kinetic theory is often criticized for its inability to predict dense granular flows. Our simulations however show very good results even in the dense regime. To be more precise, the rheology of the granular flow is investigated. Figure~\ref{fig::rheol} shows the $\mu$ versus $I$ and $\phi$ versus $I$ relations, where it is recalled that $\mu = \tau^p/p^p$ is the effective friction coefficient and $ I = d\dot{\gamma}/\sqrt{P^p/\rho^p}$ is the inertial number. The DEM data are barely dependent on the restitution coefficients and the frictional nature of particles influence strongly the rheological properties of the granular flow. The DEM data for frictional particles are superimposed for the whole range of inertial number. The $\mu(I)$ rheology for bedload transport (equations~\eqref{eq::muI} and \eqref{eq::phiI}) derived by \citet{maurin2016} (dashed-dotted line) is retrieved in the dense part of the granular flow ($10^{-2} < I < 5$). The departure between the data points and the $\mu(I)$ rheology for small $I$ (resp. large $I$) is characteristic of the quasi-static regime (resp. the dilute regime) that cannot be captured by a local rheology such as $\mu(I)$ and $\phi(I)$ rheology. The same observations can be made on the scaling between the packing fraction and the inertial number (figure~\ref{fig::rheol}b) with departure from the $\phi(I)$ law at low and large values of $I$. Both these scaling laws can also be computed in the two-fluid simulations, in which the effective friction coefficient is computed with the total stresses, \textit{i.e.} the sum of the elastic and the kinetic contributions. They are plotted in colored full lines in figure~\ref{fig::rheol}. Surprisingly, the $\mu(I)/\phi(I)$ rheology is retrieved in the dense part of the granular flow. The elastic stresses make the effective friction coefficient to asymptotically join the static friction coefficient $\mu_s$ in the limit of vanishing inertial number. At large inertial number, the two-fluid model over estimates the friction coefficient but as in the DEM, a deviation to the master curve is observed with a similar trend. The $\phi(I)$ rheology is also very well reproduced in the entire range of inertial number. The rheology obtained with the initial model, before application of the corrections proposed in section~\ref{sec::newmod}, is plotted in dashed line and highlights the role of these corrections in the rheological behavior. The dashed line shows two branches. In the dense regime, corresponding to the lower branch, the model fits with the DEM data for frictionless particles. The corrections to account for interparticle friction, including modification of $g_0$ and restitution coefficient in the dissipation law, make the granular flow less inertial improving quantitatively the results. In the dilute regime, corresponding to the higher branch, the behavior is completely different with the initial model for both scaling laws, with turning-point of the curves leading to an hysteresis effect. The improvements with the corrected model are clearly due to the modification of the kinetic contribution of the viscosity law, \textit{i.e.} $\eta_k^*$ in $F_2(\phi)$. Indeed, in the dilute regime, the elastic stresses are zero, the effective friction coefficient may be expressed as  $\mu = F_2/\sqrt{F_1}I$, and the modification of $F_2$ makes it possible to recover a qualitative behavior of the scaling law. A better understanding of the kinetic contribution of the viscosity law would certainly improve the rheological results in the dilute regime. 

In the literature, the recent attempts to model dense granular flows with the kinetic theory have used the extended kinetic theory, consisting in the introduction of a correlation length in the dissipation term $\Gamma$. Our results suggest that the reproduction of the dense granular flow regime over an erodible bed is dominated by the competition between elastic-frictional and collisional-kinetic stresses rather than the development of velocity correlation at high volume fraction \citep{jenkins2006, jenkins2007, berzi2011, berzi2020}. Here this is the hybridization of the kinetic theory with a frictional model which makes it possible to recover the correct behavior in the dense regime.

\section{Conclusions}

In this paper, the modelling of sediment transport in the collisional regime has been studied. In order to focus on the sediment phase modelling, simple fluid-particle interactions have been considered without turbulent interactions. The sediment phase is modelled with a frictional-collisional approach \citep{johnson1987}. Comparisons with coupled fluid-DEM simulations have highlighted that the classical Kinetic Theory model of \citet{garzo1999} is not able to represent correctly the behavior of the granular flow. It is shown that inter-particle friction influences the radial distribution function $g_0$ and increases energy dissipation. In the dilute regime, saltation modifies the viscosity and diffusivity laws. Finally the quadratic nature has to be taken into account because it increases the granular energy dissipation. Based on these observations, a modified KT model has been proposed that reproduces quantitatively the DEM results in the entire depth structure of the granular flow.

This work has also shown that, in the bedload configuration, the frictional-collisional approach is consistent with the $\mu(I)$ rheology and represents a first success to reproduce the dense granular flow rheology with the Kinetic Theory. The approach presented in this paper could be easily adapted to other granular configurations, in particular to any granular flows over an erodible bed, and should also be able to reproduce the rheological properties of the flow. 

The proposed modifications on the viscosity and granular temperature diffusivity laws to account for saltation are empirical and should be tested in other configurations. The success of the corrected model to reproduce particle velocity, packing fraction and rheology in the dilute regime however shows that the KT is a relevant framework to model saltation continuously. 

Finally, the proposed corrected model represents a strong basis to further study turbulence-particle interactions and their modelling in more complex configurations such as real turbulent sediment transport with coherent structures or in presence of hydraulic structures such as pipes or bridges piles.

\section*{Acknowledgments}

Most of the computations presented in this paper were performed using the GENCI infrastructure under allocations A0060107567 and A0080107567 and the GRICAD infrastructure. LEGI is member of Labex TEC21 (Investissements d'Avenir Grant Agreement ANR-11-LABX-0030) and Labex Osug@2020 (Investissements d'Avenir Grant Agreement ANR-10-LABX-0056). The authors would like to acknowledge the financial support from the Agence Nationale de la Recherche (ANR) through the project SheetFlow (ANR-18-CE01-0003).

The authors would also like to acknowledge the fruitful discussions on granular flows with P. Frey and R. Maurin.

\appendix 
\section{Definition of Favre-averaged operators and computation of the granular temperature dissipation by the drag force}\label{sec:app1}

The average operator of the two-fluid equations is presented here. Equations~\ref{eq::massFluid}, \ref{eq::massPart}, \ref{eq::momxFluid}, \ref{eq::momxPart}, \ref{eq::momzFluid} and \ref{eq::momzPart} describe the evolution of Favre-averaged quantities. It involves the ensemble averaged operator defined as
\begin{equation}
\widetilde{\gamma} = \dfrac{1}{N}\lim_{N\rightarrow\infty}\sum_{k=1}^{N}\gamma_k,
\end{equation}
where $\gamma_k$ is a given realization $k$ of a quantity $\gamma$ and $N$ the total number of realizations. The Favre-averaged is then defined as a concentration-weighted averaged and is therefore different depending on the phase considered. For a quantity $\gamma$, the Favre-averaged over the fluid and solid phases are defined respectively as
\begin{align}
\left< \gamma \right>^f = \dfrac{\widetilde{(1-\phi)\gamma}}{1-\widetilde{\phi}},\\
\left< \gamma \right>^s = \dfrac{\widetilde{\phi\gamma}}{\widetilde{\phi}}.
\end{align}
In the main text, the brackets and tildes were dropped for clarity. Considering the fluid and particle velocities, they can therefore be decomposed into a Favre-averaged and a fluctuating component. In the present configuration, the averaged velocities are non-zero only in the streamwise direction. In addition, as the models used in this paper do not consider any fluid velocity fluctuations, they are considered to be zero. The fluid and particles velocities can therefore be expressed as
\begin{align}
\label{eq::appuf}
\bm{u^f} &= \left(\left<u_x^f\right>^f, 0, 0\right),\\
\label{eq::appvp}
\bm{v^p} &= \left(\left<v_x^p\right>^s + v_x^{p\prime}, v_y^{p\prime}, v_z^{p\prime}\right).
\end{align}\\

The aim of this section is also to compute the phase-averaged dissipation of granular temperature by the drag force. Following \cite{pahtz2015} and replacing by the form of drag force expression~\eqref{eq::dragForce} and drag coefficient~\eqref{eq::dragCoef}, this term is computed as 
\begin{align}
J_{int} &= n\left<f_{Di}v_i^{p\prime}\right>^s = \dfrac{3}{4}\dfrac{\phi\rho^f(1-\phi)^{-\zeta}}{d}\left<C_D||\bm{u^f} - \bm{v^p}||(u_i^f - v_i^p)v_i^{p\prime}\right>^s,\\
\label{eq::dragApp}
&=\dfrac{3}{4}\dfrac{\phi\rho^f(1-\phi)^{-\zeta}}{d}\left[C_{D}^{\infty}\left<||\bm{u^f} - \bm{v^p}||(u_i^f - v_i^p)v_i^{p\prime}\right>^s  + \dfrac{24.4\nu^f}{d}\left<(u_i^f - v_i^p)v_i^{p\prime}\right>^s\right].
\end{align}

Let us consider the second term into the brackets of equation~\eqref{eq::dragApp}.  Replacing with the decomposition of velocities~\eqref{eq::appuf} and \eqref{eq::appvp}, it can be expressed as,
\begin{equation}
\label{eq::appvelrel}
(u_i^f - v_i^p)v_i^{p\prime} = \left(\left<u_x^f\right>^f -  \left<v_x^p\right>^s\right)v_x^{p\prime} - v_i^{p\prime}v_i^{p\prime},
\end{equation}
and applying the Favre-averaged over the solid phase, the first term vanishes and the second is three times the granular temperature,
\begin{equation}
\label{eq::appvelrel2}
\left<(u_i^f - v_i^p)v_i^{p\prime}\right>^s = -3T.
\end{equation}

Let us consider now the first term into the brackets of equation~\eqref{eq::dragApp}. The norm of the relative velocity can be expressed as
\begin{equation}
||\bm{u^f} - \bm{v^p}|| = ||\left<\bm{u^f}\right>^f - \left<\bm{v^p}\right>^s||\sqrt{1 - 2\dfrac{v_x^{p\prime}}{\left<u_x^f\right>^f -  \left<v_x^p\right>^s} + \dfrac{v_i^{p\prime}v_i^{p\prime}}{\left(\left<u_x^f\right>^f -  \left<v_x^p\right>^s\right)^2}},
\end{equation}
and performing a Taylor expansion at first order in fluctuating velocity, it becomes
\begin{equation}
||\bm{u^f} - \bm{v^p}|| = ||\left<\bm{u^f}\right>^f - \left<\bm{v^p}\right>^s||\left(1 - \dfrac{v_x^{p\prime}}{\left<u_x^f\right>^f -  \left<v_x^p\right>^s}\right) + o(v^{p\prime2}).
\end{equation}
Multiplying by equation~\eqref{eq::appvelrel} yields
\begin{equation}
||\bm{u^f} - \bm{v^p}||(u_i^f - v_i^p)v_i^{p\prime} = ||\left<\bm{u^f}\right>^f - \left<\bm{v^p}\right>^s||\left[\left(\left<u_x^f\right>^f -  \left<v_x^p\right>^s\right)v_x^{p\prime} -  v_x^{p\prime}v_x^{p\prime} - v_i^{p\prime}v_i^{p\prime} + \dfrac{v_i^{p\prime}v_i^{p\prime}v_x^{p\prime}}{\left<u_x^f\right>^f -  \left<v_x^p\right>^s}\right]+ o(v^{p\prime3}).
\end{equation}
The last term is a third order term and can therefore be neglected. Applying the Favre-averaged over the solid phase, the first term into the brackets vanishes and it becomes
\begin{equation}
\label{eq::appvelrel3}
\left<||\bm{u^f} - \bm{v^p}||(u_i^f - v_i^p)v_i^{p\prime}\right>^s = ||\left<\bm{u^f}\right>^f - \left<\bm{v^p}\right>^s||\left( -\left<v_x^{p\prime}v_x^{p\prime}\right>^s -3T \right).
\end{equation}

Replacing equations~\eqref{eq::appvelrel2} and \eqref{eq::appvelrel3} in the drag term expression~\eqref{eq::dragApp}, it is expressed as 
\begin{align}
J_{int} &= \dfrac{3}{4}\dfrac{\phi\rho^f(1-\phi)^{-\zeta}}{d} ||\left<\bm{u^f}\right>^s - \left<\bm{v^p}\right>^s|| \left[ - C_{D}^{\infty}\left<v_x^{p\prime}v_x^{p\prime}\right>^s - C_{D}^{\infty}3T - \dfrac{24.4\nu^f}{d||\left<\bm{u^f}\right>^s - \left<\bm{v^p}\right>^s||}3T    \right],\\
&= \dfrac{3}{4}\dfrac{\phi\rho^f(1-\phi)^{-\zeta}}{d} ||\left<\bm{u^f}\right>^s - \left<\bm{v^p}\right>^s||C_{D} \left(- \dfrac{C_{D}^{\infty}}{C_D}\left<v_x^{p\prime}v_x^{p\prime}\right>^s - 3T \right),
\end{align}
where $C_D$ is the drag coefficient computed with phase averaged velocities.

\begin{figure}
	\centering
	\includegraphics[width=.5\linewidth]{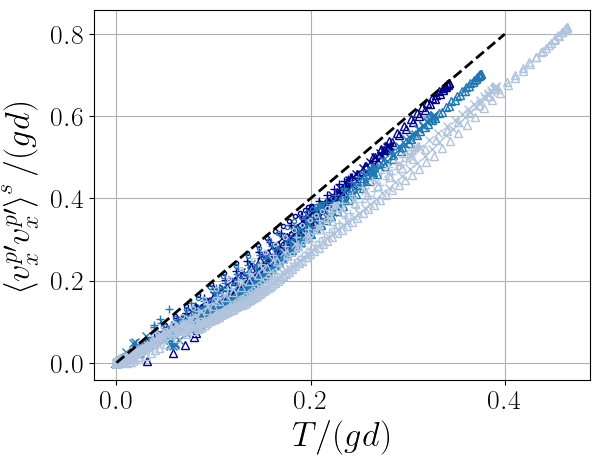}
	\caption{Streamwise fluctuating kinetic energy as a function of the granular temperature computed in the DEM simulations. Same symbols as in figure~\ref{fig::g0}. Dashed line is $2T$.}
	\label{fig::fluctuatingVel}
\end{figure}

In the bedload configuration, velocity fluctuations are not isotropic and, as shown in figure~\ref{fig::fluctuatingVel}, the streamwise fluctuating energy can be approximately estimated as $\left<v_x^{p\prime}v_x^{p\prime}\right>^s \sim 2T$. Finally the drag interaction term becomes
\begin{equation}
J_{int} = -\phi(1-\phi)K(3 + 2\dfrac{C_{D}^{\infty}}{C_{D}})T.
\end{equation}
The drag coefficient $K$ is expressed using the Favre-averaged velocities and the term $2\dfrac{C_{D}^{\infty}}{C_{D}}$ represents a correction to account for the quadratic nature of the drag force.

\bibliography{biblio}
\end{document}